\documentclass[prl,twocolumn]{revtex4-1}

\usepackage{type1cm}
\usepackage[dvipdfmx]{graphicx}

\usepackage{amsmath,amssymb,bm,amsthm}

\def\beq{\begin{equation}}
\def\eeq{\end{equation}}
\def\nbeq{\begin{equation*}}
\def\neeq{\end{equation*}}

\def\<{\langle}
\def\>{\rangle}

\renewcommand{\d}{\partial}

\begin{document}
\title{Thermodynamics of extensive but nonadditive systems: modified Gibbs-Duhem equation in the dipolar gas}

\author{Takashi Mori}
\email{
mori@spin.phys.s.u-tokyo.ac.jp}
\affiliation{
Department of Physics, Graduate School of Science,
University of Tokyo, Bunkyo-ku, Tokyo 113-0033, Japan
}

\begin{abstract}
Thermodynamic properties of extensive but nonadditive systems are investigated.
The precise definitions of additivity and extensivity are presented, and we will see that additivity derives several important properties including the shape-independence of the thermodynamic functions, the concavity of the entropy, and the equivalence of ensembles.
In nonadditive systems, some of the above properties can be violated.
It is pointed out that the shape-dependence of the entropy density in a nonadditive system results in the violation of the Gibbs-Duhem equation.
As an example, the dipolar gas is numerically studied and the violation of the Gibbs-Duhem equation is confirmed.
The predicted violation of the Gibbs-Duhem equation should be observable in cold-atom experiment with polarized dipolar gas.
\end{abstract}
\maketitle


Additivity and extensivity are central concepts in thermodynamics and statistical mechanics.
Additive and extensive systems exhibit several important properties including the concavity or convexity of thermodynamic functions and the ensemble equivalence~\cite{Campa_text,Campa_review2009,Les_Houches2009,Lecture_notes2002}.
Extensivity means that we can scale the size of the system without changing its intrinsic properties.
The system is said to be additive if any pair of macroscopic subsystems are independent of each other in thermal equilibrium. 
Short-range interacting systems with some natural conditions are additive and extensive~\cite{Ruelle_text}, and thermodynamics and statistical mechanics of those systems have been established and well understood~\cite{Ruelle_text,Landau_stat}.
On the other hand, our understanding of nonadditive systems is limited.
Long-range interacting systems are typical nonadditive systems, and they exhibit interesting phenomenologies~\cite{Campa_text,Campa_review2009,Les_Houches2009,Lecture_notes2002}.
Furthermore, recent experimental progress has made long-range interacting systems accessible in the laboratory, and it becomes possible to explore their properties~\cite{Britton2012,Islam2013,Chalony2013,Senko2015}.
It is desired to formulate thermodynamics of nonadditive systems.

We should carefully distinguish the extensivity and additivity~\cite{Campa_text,Campa_review2009}.
Although long-range interacting systems are in general neither extensive nor additive, the extensivity can be restored by introducing the size-dependent scaling to the interaction potential, which is called Kac prescription~\cite{Kac1963}.
It is noted that the additivity is not recovered by such a scaling.
Moreover, systems with dipole-dipole interactions, which are ubiquitous in nature, are considered to be extensive but nonadditive even without Kac prescription, as discussed later.
Thus, additivity and extensivity are different concepts, and in order to focus on the consequences of nonadditivity, we consider extensive but nonadditive systems in this paper.

The aim of this paper is to formulate thermodynamics of extensive but nonadditive systems and investigate its characteristics.
We first start from the precise definition of extensivity and additivity because these terminologies tend to be used ambiguously.
As was mentioned above, additivity roughly means the independence of macroscopic subsystems in thermal equilibrium.
Problem here is that the notion of the independence is a subtle concept.
Usually, the two macroscopic subsystems are considered to be independent if the interaction energy between them is negligible compared to the bulk energy.
However, it has been recognized that in some models two macroscopic subsystems are strongly correlated although the interaction energy is extremely small compared to the bulk energy~\cite{Mori2013_nonadditivity,Mori2015_quasi}.
We give a precise definition of additivity without ambiguity, and then investigate thermodynamic properties of nonadditive systems.
Suitability of the definition presented here will be confirmed by showing that this additivity automatically derives several important and natural properties such as the shape independence of the thermodynamic functions, their concavity or convexity, and the ensemble equivalence.

Recently, by formulating thermodynamics of nonadditive and/or nonextensive systems following Hill's formulation of thermodynamics of small systems~\cite{Hill_text}, Latella et al.~\cite{Latella2015} argued that the Gibbs-Duhem equation $SdT-VdP+Nd\mu=0$ should be modified in such systems, where $U$ is the internal energy, $V$ the volume, $N$ the number of particles, $T$ the temperature, $P$ the pressure and $\mu$ the chemical potential.
The modified equation reads $SdT-VdP+Nd\mu=-d\mathcal{E}$, where $\mathcal{E}$ is called the ``replica energy''~\cite{Hill_text,Latella2015,Latella2013}.
In Ref.~\cite{Latella2015} the replica energy is explicitly computed only for classical particle systems with purely long-range interactions.
However, such systems have no thermodynamic limit and are therefore \textit{nonextensive} even if we apply the Kac prescription because sufficiently strong short-range repulsions are necessary for the existence of the thermodynamic limit~\cite{Ruelle_text}.
Therefore, it has not been solved whether and how the Gibbs-Duhem equation should be modified in extensive but nonadditive systems.
Here, we show that the shape dependence of the equilibrium state plays a significant role in thermodynamics of extensive but nonadditive systems, and the Gibbs-Duhem equation can be violated owing to this shape dependence.

Let us start with presenting the definition of additivity and extensivity.
Suppose a macroscopic system enclosed in the domain $\Gamma \subset \mathbb{R}^d$ of volume $|\Gamma|=V$ with the internal energy $U=Vu$ and the number of particles $N=V\rho$.
Here, $d$ is the spacial dimension.
Because the domain $\Gamma$ is characterized by its volume $V$ and its shape denoted by $\gamma$,
let us write the entropy as $S(U,V,N;\gamma)$.
For example, we can specify the shape of the system by putting $\gamma=\Gamma/V^{1/d}$, where we define $aX\equiv\{\bm{x}\in\mathbb{R}^d:\bm{x}/a\in X\}$ for a domain $X\subset\mathbb{R}^d$ and a positive value $a>0$; $\gamma$ is the normalized domain such that the volume of $\gamma$ is unity, $|\gamma|=1$.

The definition of extensivity is simple.
Let us change the system size with the energy density and the particle density held fixed, that is, $\Gamma\rightarrow \alpha^{1/d}\Gamma$ ($V\rightarrow \alpha V$), $U\rightarrow \alpha U$, and $N\rightarrow \alpha N$.
If the entropy changes as $S\rightarrow \alpha S$ in this transformation, the system is said to be extensive.
More precisely, the system is said to be extensive if
\beq
S(U,V,N;\gamma)\approx Vs_{\gamma}(u,\rho)
\eeq
for sufficiently large systems, where $s_{\gamma}$ is the entropy density independent of the volume.
Essentially this is equivalent to the statement that the system has a well-defined thermodynamic limit for each shape $\gamma$.

Next, we consider the definition of additivity.
Let us consider the system in the domain $\Gamma\subset\mathbb{R}^d$ with $|\Gamma|=V$ consisting of the two subsystems $A$ and $B$.
The domains of $A$ and $B$ are respectively $\Gamma_A$ and $\Gamma_B$ and we define the normalized domain $\gamma_A\equiv\Gamma_A/V^{1/d}$ and $\gamma_B\equiv\Gamma_B/V^{1/d}$ in order to specify the shape of the subsystems.
Note that $\gamma_A\cap\gamma_B=\emptyset$ and $|\gamma_A\cup\gamma_B|=1$.
Here we define $\lambda$ as $|\gamma_A|=\lambda$ and $|\gamma_B|=1-\lambda$.
The Hamiltonian is given by $H=H_A+H_B+\eta H_{\rm int}$, where $H_A (H_B)$ is the Hamiltonian of the system $A (B)$ when there is no interaction between $A$ and $B$.
The interaction energy is $\eta H_{\rm int}$ and one might be tempted to consider that the system is additive if $\eta H_{\rm int}$ is very small compared to the total energy.
However, in some cases this naive definition does not work; there is a physical model in which the influence of the interaction term is strong although the typical value of $\eta H_{\rm int}$ is negligibly small~\cite{Mori2013_nonadditivity,Mori2015_quasi}.

The smallness of the influence of the interaction term can be formulated in the following way.
Let us consider the situation in which the particle exchange between $A$ and $B$ is prohibited, while the energy exchange is allowed.
In that case, the state of the total system is specified by $(u,\rho_A,\rho_B)$, where $u$ is the total energy density and $\rho_A (\rho_B)$ is the particle density of the subsystem $A (B)$.
We consider a quasi-static adiabatic process to decouple the two subsystems in this situation, that is, decrease $\eta$ to zero very slowly with $\rho_A$ and $\rho_B$ held constant.
The amount of work done by the system during this thermodynamic process is denoted by $W=V(u-u')$, where $u$ and $u'$ are the energy densities before and after the thermodynamic process, respectively.
We now define additivity from the thermodynamic point of view.
\textit{If $W/V$ is negligibly small for large $V$, i.e. if we can divide the two subsystems without performing work, we say that the system is additive}.

From now on, let us see the consequence of the additivity.
The entropy of the final state is written as
\beq
s'=\sup_{\substack{u_A,u_B \\ \lambda u_A+(1-\lambda)u_B=u}}\left[\lambda s_{\gamma_A}(u_A,\rho_A)+(1-\lambda)s_{\gamma_B}(u_B,\rho_B)\right],
\eeq
when the system is additive (otherwise, the relation $\lambda u_A+(1-\lambda)u_B=u$ in the right-hand side should be replaced by $\lambda u_A+(1-\lambda)u_B=u'$).
Since we consider the quasi-static adiabatic process, the entropy is invariant, and hence the entropy density of the initial state, i.e., the equilibrium state with $(u,\rho_A,\rho_B)$, is identical to $s'$. 
Thus the entropy density of the single system in the domain $\Gamma_A\cup\Gamma_B$, in which both the energy and the particle exchange are allowed between $A$ and $B$, is obtained by maximizing $s'$ over $\rho_A$ and $\rho_B$ under the restriction $\lambda\rho_A+(1-\lambda)\rho_B=\rho$, where $\rho$ is the particle density of the whole system:
\begin{align}
s_{\gamma_A\cup\gamma_B}(u,\rho)=\sup_{\substack{u_A,u_B \\ \lambda u_A+(1-\lambda)u_B=u}} \sup_{\substack{\rho_A,\rho_B \\ \lambda\rho_A+(1-\lambda)\rho_B=\rho}}
\nonumber \\
\left[\lambda s_{\gamma_A}(u_A,\rho_A)+(1-\lambda)s_{\gamma_B}(u_B,\rho_B)\right].
\label{eq:entropy}
\end{align}
From this equation, we find that the entropy density of an additive system does not depend on the shape of the system because the right-hand side is independent of the relative position of $\gamma_A$ and $\gamma_B$.
We can therefore simply write $s_{\gamma}(u,\rho)=s(u,\rho)$.
Moreover, eq.~(\ref{eq:entropy}) implies
\begin{align}
s(\lambda u_A+(1-\lambda)u_B,\lambda\rho_A+(1-\lambda)\rho_B)
\nonumber \\
\geq \lambda s(u_A,\rho_A)+(1-\lambda)s(u_B,\rho_B),
\end{align}
which shows the concavity of the entropy density.
From the concavity of the entropy, we can show the equivalence of the microcanonical, canonical, and grandcanonical ensembles~\cite{Mori2013_statphys}.
The ensemble equivalence also ensures that the specific heat is always nonnegative in the microcanonical ensemble.

In this way, starting from the definition of additivity, we can show several important results.
Thus the additivity defined in the above argument is considered to be a fundamental property.

As was mentioned, additive systems {\it always} have a shape-independent entropy density in the thermodynamic limit.
An important characteristic of nonadditive systems is thus the shape dependence of the thermodynamic functions.
In Ref.~\cite{Mori2015_existence}, existence of the shape-dependent thermodynamic limit is proved for long-range interacting lattice systems under the Kac prescription.
Such systems are therefore extensive but not additive.

Electric or magnetic dipolar systems are also extensive but nonadditive.
Griffiths~\cite{Griffiths1968} proved that a spin system with magnetic dipole-dipole interactions possesses a shape-independent thermodynamic limit if there is no external magnetic field (the proof for off-lattice particles is given in Ref.~\cite{Banerjee1998}).
On the other hand, it is widely believed that the thermodynamic functions depend on the shape of the system even in the thermodynamic limit when the external magnetic field is present~\cite{Sauer1940,Hiley-Joyce1965,Levy1968,Horner1968}.
The shape dependence of the thermodynamic limit means that a dipolar system is not additive although it is extensive.


Because of the shape-dependence of thermodynamic quantities, the first-law of thermodynamics should be modified as
\beq
dS=\frac{dU}{T_{\gamma}}+\frac{P_{\gamma}}{T_{\gamma}}dV-\frac{\mu_{\gamma}}{T_{\gamma}}dN-\frac{K}{T_{\gamma}}d\gamma,
\label{eq:1st}
\eeq
where subscript $\gamma$ means that the quantities are defined for a fixed shape $\gamma$.
Here, the work done in an infinitesimal adiabatic process in which the shape of the system changes by $d\gamma$ is symbolically denoted by $Kd\gamma$~\footnote{Because the shape $\gamma$ can be characterized by a function giving the surface of $\gamma$, $Kd\gamma$ can be expressed as some functional derivative of the entropy.}.

By using the fact that the system is extensive, $\alpha S(U,V,N;\gamma)=S(\alpha U,\alpha V,\alpha N;\gamma)$ with $\alpha>0$, we can derive
\beq
SdT_{\gamma}-VdP_{\gamma}+Nd\mu_{\gamma}=Kd\gamma.
\label{eq:modified_GD}
\eeq
If the shape of the system is held fixed, $d\gamma=0$, eq.~(\ref{eq:modified_GD}) reduces to the usual Gibbs-Duhem equation.
When the system has a fixed volume $V$ but can change its shape spontaneously without work, the shape $\gamma$ is determined by the condition of the maximum entropy.
This implies $K=0$ and also in this case, eq.~(\ref{eq:modified_GD}) reduces to the usual Gibbs-Duhem equation.
Thus extensive but nonadditive systems satisfy the Gibbs-Duhem equation in these situations.

\begin{figure}[t]
\begin{center}
\includegraphics[clip,width=8cm]{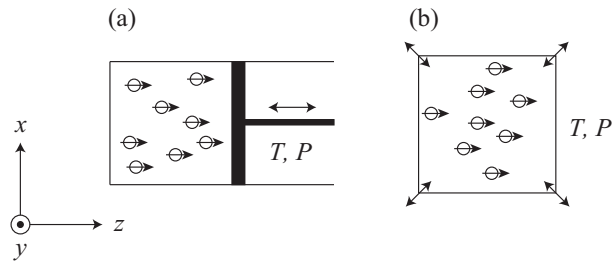}
\caption{The dipolar particles are confined in a box surrounded by the environment with the temperature $T$ and the pressure $P$. All the dipoles are oriented to $z$-direction. 
(a) The volume of the system is controlled by the piston.
(b) The shape of the system is fixed to be cubic.}
\label{fig:setup}
\end{center}
\end{figure}

However, in general, the term $(K/T_{\gamma})d\gamma$ in eq.~(\ref{eq:1st}) or $Kd\gamma$ in eq.~(\ref{eq:modified_GD}) plays a role.
As an example, suppose the situation in which $N$ particles are enclosed in the container of volume $V$ and $V$ can be controlled by moving the piston, see Fig.~\ref{fig:setup} (a).
In this case, the shape of the system $\gamma$ changes by moving the piston, and hence $\gamma$ should be considered as a function of $V$.
Therefore, $d\gamma=(d\gamma/dV)dV$ and eq.~(\ref{eq:1st}) becomes
\beq
dS=\frac{dU}{T_{\gamma}}+\frac{1}{T_{\gamma}}\left(P_{\gamma}-K\frac{d\gamma}{dV}\right)dV-\frac{\mu_{\gamma}}{T_{\gamma}}dN.
\label{eq:1st_piston}
\eeq
The shape of the system is no longer an independent variable in this situation and the usual first-law of thermodynamics $dS=(dU+PdV-\mu dN)/T$ is applicable. Therefore $T=T_{\gamma}$, $\mu=\mu_{\gamma}$, and
\beq
P=P_{\gamma}-K\frac{d\gamma}{dV},
\label{eq:pressure}
\eeq
where $P$ is the physical pressure in this situation (the work done by moving the piston infinitesimally is given by $-PdV$), while $-P_{\gamma}dV$ is the amount of work done in an adiabatic process in which the volume $V$ changes to $V+dV$ \textit{keeping the shape $\gamma$} as in Fig.~\ref{fig:setup} (b), that is, $P_{\gamma}=-(\d U/\d V)_{S,N,\gamma}$.
Substituting eq.~(\ref{eq:pressure}) into eq.~(\ref{eq:modified_GD}), we obtain the result
\beq
SdT-VdP+Nd\mu=d\left(VK\frac{d\gamma}{dV}\right).
\label{eq:GD_violation}
\eeq
This result is consistent with the formulation given in Ref.~\cite{Latella2015}, in which the violation of the Gibbs-Duhem equation is given by the total derivative of the replica energy $-d\mathcal{E}$.
We can identify $\mathcal{E}=-VKd\gamma/dV$ in this example.

As was already pointed out in Ref.~\cite{Latella2013}, the nonadditivity introduces an extra degree of freedom $\mathcal{E}$.
In this paper, it has been clarified that this extra degree of freedom originates from the shape dependence of the thermodynamic functions. 
It is emphasized that this extra degree of freedom appears even if the shape $\gamma$ is \textit{not} an independent variable; in eq.~(\ref{eq:GD_violation}), $\gamma$ is a function of $V$ and independent variables are just $U$, $V$, and $N$.

In order to numerically confirm the violation of the Gibbs-Duhem equation, we consider classical particles with the dipole-dipole interactions and the hard-core repulsions in the container with the piston, see Fig.~\ref{fig:setup} (a).
We assume that very strong magnetic field is applied to the $z$-direction, and hence all the magnetic dipoles are oriented to the $z$-direction.
The pressure $P$ is applied to the piston and the system is in contact with a thermal bath at the temperature $T$.
The Hamiltonian of our three-dimensional system is written as
\beq
H=\sum_{i=1}^N\frac{\bm{p}_i^2}{2m}+\sum_{i<j}V(\bm{r}_i-\bm{r}_j),
\label{eq:Ham}
\eeq
where the position and the momentum of $i$th particle are denoted by $\bm{r}_i$ and $\bm{p}_i$, respectively, and the mass of the particle is denoted by $m$ (in numerical calculations, we put $m=1$).
The dipole-dipole interaction reads
\beq
V(\bm{r}_{ij})=
\left\{
\begin{aligned}
&\frac{1-3\cos^2\theta_{ij}}{r_{ij}^3} &\text{for } r_{ij}>a, \\
&+\infty &\text{otherwise},
\end{aligned}
\right.
\eeq
where $\bm{r}_{ij}=\bm{r}_i-\bm{r}_j$ and $\theta_{ij}$ is the angle between $\bm{r}_{ij}$ and the $z$-direction.

Since the usual form of the Gibbs-Duhem equation contains the entropy which is hard to calculate numerically, we consider the Gibbs-Duhem equation in the entropy representation, $Ud\beta+Vd(\beta P)-Nd(\beta\mu)=0$~\cite{Callen_text}, where $\beta=1/T$ (Boltzmann constant is put unity).
When the system is not additive, this equation is also modified and the equation corresponding to eq.~(\ref{eq:GD_violation}) reads
\beq
Ud\beta+Vd(\beta P)-Nd(\beta\mu)=-d\left(\beta VK\frac{d\gamma}{dV}\right).
\label{eq:GD_entropy}
\eeq 

\begin{figure}[t]
\begin{center}
\includegraphics[clip,width=6cm]{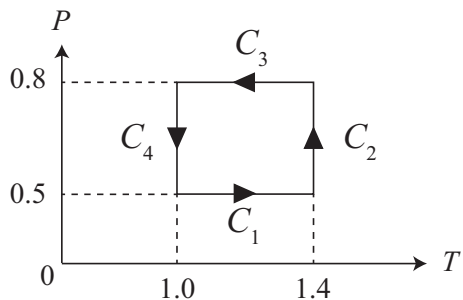}
\caption{Paths $\{C_i\}_{i=1}^4$ of the thermodynamic process.
During the thermodynamic process, the number of particles $N$ is held fixed.}
\label{fig:paths}
\end{center}
\end{figure}

\begin{table}[t]
\begin{tabular}{|l|c|rrrrr|}
\hline
&$N$ & $\Delta_1$ & $\Delta_2$ & $\Delta_3$ & $\Delta_4$ & $\sum_{i=1}^4\Delta_i$ \\ \hline
(a)&50 & $-10.60(3)$ & 3.71(2) & 10.08(3) & $-3.07(3)$ & 0.12(6) \\ \cline{2-7}
piston& 100 & $-21.1(1)$ & 7.7(1) & 22.1(1) & $-8.3(2)$ & 0.4(3) \\ \cline{2-7}
& 200 & $-43.0(4)$ & 16.6(1) & 45.3(5) & $-20.3(4)$ & $-1.4(7)$ \\ \hline\hline
(b)& 50 & $-1.29(3)$ & 0.89(2) & 1.24(3) & $-0.71(2)$ & 0.13(5) \\ \cline{2-7}
fixed & 100 & $-1.0(1)$ & 0.3(1) & 0.5(1) & 0.2(2) & $-0.1(3)$ \\ \cline{2-7}
shape & 200 & $-0.5(2)$ & $-1.4(1)$ & $-0.5(2)$ & $1.8(5)$ & $-0.6(6)$ \\ \hline
\end{tabular}
\caption{Numerical data of $\Delta_i$ calculated by the Monte-Carlo method.
Each path is divided into 72 segments to calculate the integration.
(a) The data for the case of Fig.~\ref{fig:setup} (a), in which the volume is controlled by moving the piston.
(b) The data for the case of Fig.~\ref{fig:setup} (b), in which the cubic shape of the system is held fixed.}
\label{table}
\end{table}

For a given set of $(T,P,N)$, we numerically evaluate $(U,V,\mu)$ by the Monte-Carlo method~\cite{Landau-Binder_text}.
Then we calculate 
\beq
\Delta_i\equiv\int_{C_i}[Ud\beta+Vd(\beta P)-Nd(\beta\mu)]
\eeq
along the paths $C_1$, $C_2$, $C_3$, and $C_4$ given in Fig.~\ref{fig:paths}.
Note that the whole path $\{ C_i\}_{i=1}^4$ forms the closed loop.
The numerical result is given in Table~\ref{table}.

When we consider the situation of Fig.~\ref{fig:setup} (a), in which the volume varies by moving the piston, Table~\ref{table} (a) clearly shows that the usual Gibbs-Duhem equation is violated.
For each $i$, $\Delta_i$ is proportional to the system size.
On the other hand, the integration along the closed loop, $\sum_{i=1}^4\Delta_i$, remains small, which is consistent with the fact that the degree of the violation of the Gibbs-Duhem equation is written by a total derivative as in eq.~(\ref{eq:GD_entropy}).

On the other hand, when the shape of the system is fixed (suppose the cubic shape) and the system is made large or small uniformly as in Fig.~\ref{fig:setup} (b), the Gibbs-Duhem equation should be satisfied, as discussed below eq.~(\ref{eq:modified_GD}).
We also calculated $\{\Delta_i\}_{i=1}^4$ numerically for this case and the result is presented in Table~\ref{table} (b).
It is found that $\Delta_i$ does not grow as the system size increases, which means that the Gibbs-Duhem equation holds.
Thus the Gibbs-Duhem equation is not always violated in extensive but nonadditive systems.

Systems with significant dipole-dipole interactions can be realized in cold atoms~\cite{Lahaye2009review}.
Therefore, the violation of the Gibbs-Duhem equation should be observable in a polarized dipolar gas.
For a harmonically trapped gas with the trap frequencies $\omega_x$, $\omega_y$, and $\omega_z$, the usual Gibbs-Duhem equation should hold if we change $\omega_x$, $\omega_y$, and $\omega_z$ uniformly to control the volume of the system (see Ref.~\cite{Sandoval-Figueroa2008} for the definition of the volume and the pressure for trapped systems).
On the other hand, if we change only one of them, say $\omega_z$, the situation is essentially the same as that in Fig.~\ref{fig:setup} (a), and the Gibbs-Duhem equation should be modified.

In conclusion, thermodynamics of extensive but nonadditive systems have been formulated.
As a remarkable property, it has been shown that the Gibbs-Duhem equation can be violated in such systems.
This violation stems from the shape dependence of the thermodynamic functions.
The degree of the violation strongly depends on the type of the macroscopic operations done to the system, e.g., how to control the volume of the system.
This theoretical prediction has been confirmed for classical dipolar particles with the hard-core repulsion.
It should be also verified in experiment for polarized dipolar gas.
Since the violation of the Gibbs-Duhem equation results from the general thermodynamic formalism, it should be a universal property in nonadditive systems; the quantum effect, for instance, will not alter this property although the numerical model studied here is a classical one.

The author thanks Ivan Latella for stimulating discussions and Seiji Miyashita for useful comments.
This work was supported by the Elements Strategy Initiative Center for Magnetic Materials under the outsourcing project of MEXT.


\begin{thebibliography}{29}%
\makeatletter
\providecommand \@ifxundefined [1]{%
 \@ifx{#1\undefined}
}%
\providecommand \@ifnum [1]{%
 \ifnum #1\expandafter \@firstoftwo
 \else \expandafter \@secondoftwo
 \fi
}%
\providecommand \@ifx [1]{%
 \ifx #1\expandafter \@firstoftwo
 \else \expandafter \@secondoftwo
 \fi
}%
\providecommand \natexlab [1]{#1}%
\providecommand \enquote  [1]{``#1''}%
\providecommand \bibnamefont  [1]{#1}%
\providecommand \bibfnamefont [1]{#1}%
\providecommand \citenamefont [1]{#1}%
\providecommand \href@noop [0]{\@secondoftwo}%
\providecommand \href [0]{\begingroup \@sanitize@url \@href}%
\providecommand \@href[1]{\@@startlink{#1}\@@href}%
\providecommand \@@href[1]{\endgroup#1\@@endlink}%
\providecommand \@sanitize@url [0]{\catcode `\\12\catcode `\$12\catcode
  `\&12\catcode `\#12\catcode `\^12\catcode `\_12\catcode `\%12\relax}%
\providecommand \@@startlink[1]{}%
\providecommand \@@endlink[0]{}%
\providecommand \url  [0]{\begingroup\@sanitize@url \@url }%
\providecommand \@url [1]{\endgroup\@href {#1}{\urlprefix }}%
\providecommand \urlprefix  [0]{URL }%
\providecommand \Eprint [0]{\href }%
\providecommand \doibase [0]{http://dx.doi.org/}%
\providecommand \selectlanguage [0]{\@gobble}%
\providecommand \bibinfo  [0]{\@secondoftwo}%
\providecommand \bibfield  [0]{\@secondoftwo}%
\providecommand \translation [1]{[#1]}%
\providecommand \BibitemOpen [0]{}%
\providecommand \bibitemStop [0]{}%
\providecommand \bibitemNoStop [0]{.\EOS\space}%
\providecommand \EOS [0]{\spacefactor3000\relax}%
\providecommand \BibitemShut  [1]{\csname bibitem#1\endcsname}%
\let\auto@bib@innerbib\@empty
\bibitem [{\citenamefont {Campa}\ \emph {et~al.}(2014)\citenamefont {Campa},
  \citenamefont {Dauxois}, \citenamefont {Fanelli},\ and\ \citenamefont
  {Ruffo}}]{Campa_text}%
  \BibitemOpen
  \bibfield  {author} {\bibinfo {author} {\bibfnamefont {A.}~\bibnamefont
  {Campa}}, \bibinfo {author} {\bibfnamefont {T.}~\bibnamefont {Dauxois}},
  \bibinfo {author} {\bibfnamefont {D.}~\bibnamefont {Fanelli}}, \ and\
  \bibinfo {author} {\bibfnamefont {S.}~\bibnamefont {Ruffo}},\ }\href@noop {}
  {\emph {\bibinfo {title} {Physics of long-range interacting systems}}}\
  (\bibinfo  {publisher} {Oxford University Press},\ \bibinfo {year}
  {2014})\BibitemShut {NoStop}%
\bibitem [{\citenamefont {Campa}\ \emph {et~al.}(2009)\citenamefont {Campa},
  \citenamefont {Dauxois},\ and\ \citenamefont {Ruffo}}]{Campa_review2009}%
  \BibitemOpen
  \bibfield  {author} {\bibinfo {author} {\bibfnamefont {A.}~\bibnamefont
  {Campa}}, \bibinfo {author} {\bibfnamefont {T.}~\bibnamefont {Dauxois}}, \
  and\ \bibinfo {author} {\bibfnamefont {S.}~\bibnamefont {Ruffo}},\
  }\href@noop {} {\bibfield  {journal} {\bibinfo  {journal} {Phys. Rep.}\
  }\textbf {\bibinfo {volume} {480}},\ \bibinfo {pages} {57} (\bibinfo {year}
  {2009})}\BibitemShut {NoStop}%
\bibitem [{\citenamefont {Dauxois}\ \emph {et~al.}(2009)\citenamefont
  {Dauxois}, \citenamefont {Ruffo},\ and\ \citenamefont
  {Cugliandolo}}]{Les_Houches2009}%
  \BibitemOpen
  \bibfield  {author} {\bibinfo {author} {\bibfnamefont {T.}~\bibnamefont
  {Dauxois}}, \bibinfo {author} {\bibfnamefont {S.}~\bibnamefont {Ruffo}}, \
  and\ \bibinfo {author} {\bibfnamefont {L.}~\bibnamefont {Cugliandolo}},\
  }\href@noop {} {\emph {\bibinfo {title} {Long-Range Interacting Systems, vol.
  90 of Lecture Notes of the Les Houches Summer School}}}\ (\bibinfo
  {publisher} {Oxford University Press},\ \bibinfo {year} {2009})\BibitemShut
  {NoStop}%
\bibitem [{\citenamefont {Dauxois}\ \emph {et~al.}(2002)\citenamefont
  {Dauxois}, \citenamefont {Ruffo}, \citenamefont {Arimondo},\ and\
  \citenamefont {Wilkens}}]{Lecture_notes2002}%
  \BibitemOpen
  \bibfield  {author} {\bibinfo {author} {\bibfnamefont {T.}~\bibnamefont
  {Dauxois}}, \bibinfo {author} {\bibfnamefont {S.}~\bibnamefont {Ruffo}},
  \bibinfo {author} {\bibfnamefont {E.}~\bibnamefont {Arimondo}}, \ and\
  \bibinfo {author} {\bibfnamefont {M.}~\bibnamefont {Wilkens}},\ }\href@noop
  {} {\emph {\bibinfo {title} {Dynamics and Thermodynamics of Systems with Long
  Range Interactions, Lecture Notes in Physics vol 602}}}\ (\bibinfo
  {publisher} {Springer, Berlin},\ \bibinfo {year} {2002})\BibitemShut
  {NoStop}%
\bibitem [{\citenamefont {Ruelle}(1999)}]{Ruelle_text}%
  \BibitemOpen
  \bibfield  {author} {\bibinfo {author} {\bibfnamefont {D.}~\bibnamefont
  {Ruelle}},\ }\href@noop {} {\emph {\bibinfo {title} {Statistical Mechanics:
  Rigorous Results}}}\ (\bibinfo  {publisher} {World Scientific Publishing
  Company Incorporated},\ \bibinfo {year} {1999})\BibitemShut {NoStop}%
\bibitem [{\citenamefont {Landau}\ and\ \citenamefont
  {Lifshitz}(1980)}]{Landau_stat}%
  \BibitemOpen
  \bibfield  {author} {\bibinfo {author} {\bibfnamefont {L.~D.}\ \bibnamefont
  {Landau}}\ and\ \bibinfo {author} {\bibfnamefont {E.~M.}\ \bibnamefont
  {Lifshitz}},\ }\href@noop {} {\emph {\bibinfo {title} {Statistical Physics,
  Pt. 1}}}\ (\bibinfo  {publisher} {Butterworth-Heinemann, Oxford},\ \bibinfo
  {year} {1980})\BibitemShut {NoStop}%
\bibitem [{\citenamefont {Britton}\ \emph {et~al.}(2012)\citenamefont
  {Britton}, \citenamefont {Sawyer}, \citenamefont {Keith}, \citenamefont
  {Wang}, \citenamefont {Freericks}, \citenamefont {Uys}, \citenamefont
  {Biercuk},\ and\ \citenamefont {Bollinger}}]{Britton2012}%
  \BibitemOpen
  \bibfield  {author} {\bibinfo {author} {\bibfnamefont {J.~W.}\ \bibnamefont
  {Britton}}, \bibinfo {author} {\bibfnamefont {B.~C.}\ \bibnamefont {Sawyer}},
  \bibinfo {author} {\bibfnamefont {A.~C.}\ \bibnamefont {Keith}}, \bibinfo
  {author} {\bibfnamefont {C.-C.~J.}\ \bibnamefont {Wang}}, \bibinfo {author}
  {\bibfnamefont {J.~K.}\ \bibnamefont {Freericks}}, \bibinfo {author}
  {\bibfnamefont {H.}~\bibnamefont {Uys}}, \bibinfo {author} {\bibfnamefont
  {M.~J.}\ \bibnamefont {Biercuk}}, \ and\ \bibinfo {author} {\bibfnamefont
  {J.~J.}\ \bibnamefont {Bollinger}},\ }\href@noop {} {\bibfield  {journal}
  {\bibinfo  {journal} {Nature}\ }\textbf {\bibinfo {volume} {484}},\ \bibinfo
  {pages} {489} (\bibinfo {year} {2012})}\BibitemShut {NoStop}%
\bibitem [{\citenamefont {Islam}\ \emph {et~al.}(2013)\citenamefont {Islam},
  \citenamefont {Senko}, \citenamefont {Campbell}, \citenamefont {Korenblit},
  \citenamefont {Smith}, \citenamefont {Lee}, \citenamefont {Edwards},
  \citenamefont {Wang}, \citenamefont {Freericks},\ and\ \citenamefont
  {Monroe}}]{Islam2013}%
  \BibitemOpen
  \bibfield  {author} {\bibinfo {author} {\bibfnamefont {R.}~\bibnamefont
  {Islam}}, \bibinfo {author} {\bibfnamefont {C.}~\bibnamefont {Senko}},
  \bibinfo {author} {\bibfnamefont {W.}~\bibnamefont {Campbell}}, \bibinfo
  {author} {\bibfnamefont {S.}~\bibnamefont {Korenblit}}, \bibinfo {author}
  {\bibfnamefont {J.}~\bibnamefont {Smith}}, \bibinfo {author} {\bibfnamefont
  {A.}~\bibnamefont {Lee}}, \bibinfo {author} {\bibfnamefont {E.}~\bibnamefont
  {Edwards}}, \bibinfo {author} {\bibfnamefont {C.-C.}\ \bibnamefont {Wang}},
  \bibinfo {author} {\bibfnamefont {J.}~\bibnamefont {Freericks}}, \ and\
  \bibinfo {author} {\bibfnamefont {C.}~\bibnamefont {Monroe}},\ }\href@noop {}
  {\bibfield  {journal} {\bibinfo  {journal} {Science}\ }\textbf {\bibinfo
  {volume} {340}},\ \bibinfo {pages} {583} (\bibinfo {year}
  {2013})}\BibitemShut {NoStop}%
\bibitem [{\citenamefont {Chalony}\ \emph {et~al.}(2013)\citenamefont
  {Chalony}, \citenamefont {Barr\'e}, \citenamefont {Marcos}, \citenamefont
  {Olivetti},\ and\ \citenamefont {Wilkowski}}]{Chalony2013}%
  \BibitemOpen
  \bibfield  {author} {\bibinfo {author} {\bibfnamefont {M.}~\bibnamefont
  {Chalony}}, \bibinfo {author} {\bibfnamefont {J.}~\bibnamefont {Barr\'e}},
  \bibinfo {author} {\bibfnamefont {B.}~\bibnamefont {Marcos}}, \bibinfo
  {author} {\bibfnamefont {A.}~\bibnamefont {Olivetti}}, \ and\ \bibinfo
  {author} {\bibfnamefont {D.}~\bibnamefont {Wilkowski}},\ }\href {\doibase
  10.1103/PhysRevA.87.013401} {\bibfield  {journal} {\bibinfo  {journal} {Phys.
  Rev. A}\ }\textbf {\bibinfo {volume} {87}},\ \bibinfo {pages} {013401}
  (\bibinfo {year} {2013})}\BibitemShut {NoStop}%
\bibitem [{\citenamefont {Senko}\ \emph {et~al.}(2015)\citenamefont {Senko},
  \citenamefont {Richerme}, \citenamefont {Smith}, \citenamefont {Lee},
  \citenamefont {Cohen}, \citenamefont {Retzker},\ and\ \citenamefont
  {Monroe}}]{Senko2015}%
  \BibitemOpen
  \bibfield  {author} {\bibinfo {author} {\bibfnamefont {C.}~\bibnamefont
  {Senko}}, \bibinfo {author} {\bibfnamefont {P.}~\bibnamefont {Richerme}},
  \bibinfo {author} {\bibfnamefont {J.}~\bibnamefont {Smith}}, \bibinfo
  {author} {\bibfnamefont {A.}~\bibnamefont {Lee}}, \bibinfo {author}
  {\bibfnamefont {I.}~\bibnamefont {Cohen}}, \bibinfo {author} {\bibfnamefont
  {A.}~\bibnamefont {Retzker}}, \ and\ \bibinfo {author} {\bibfnamefont
  {C.}~\bibnamefont {Monroe}},\ }\href {\doibase 10.1103/PhysRevX.5.021026}
  {\bibfield  {journal} {\bibinfo  {journal} {Phys. Rev. X}\ }\textbf {\bibinfo
  {volume} {5}},\ \bibinfo {pages} {021026} (\bibinfo {year}
  {2015})}\BibitemShut {NoStop}%
\bibitem [{\citenamefont {Kac}\ \emph {et~al.}(1963)\citenamefont {Kac},
  \citenamefont {Uhlenbeck},\ and\ \citenamefont {Hemmer}}]{Kac1963}%
  \BibitemOpen
  \bibfield  {author} {\bibinfo {author} {\bibfnamefont {M.}~\bibnamefont
  {Kac}}, \bibinfo {author} {\bibfnamefont {G.~E.}\ \bibnamefont {Uhlenbeck}},
  \ and\ \bibinfo {author} {\bibfnamefont {P.~C.}\ \bibnamefont {Hemmer}},\
  }\href@noop {} {\bibfield  {journal} {\bibinfo  {journal} {J. Math. Phys.}\
  }\textbf {\bibinfo {volume} {4}},\ \bibinfo {pages} {216} (\bibinfo {year}
  {1963})}\BibitemShut {NoStop}%
\bibitem [{\citenamefont {Mori}(2013{\natexlab{a}})}]{Mori2013_nonadditivity}%
  \BibitemOpen
  \bibfield  {author} {\bibinfo {author} {\bibfnamefont {T.}~\bibnamefont
  {Mori}},\ }\href@noop {} {\bibfield  {journal} {\bibinfo  {journal} {Phys.
  Rev. Lett.}\ }\textbf {\bibinfo {volume} {111}},\ \bibinfo {pages} {020601}
  (\bibinfo {year} {2013}{\natexlab{a}})}\BibitemShut {NoStop}%
\bibitem [{\citenamefont {Mori}(2015{\natexlab{a}})}]{Mori2015_quasi}%
  \BibitemOpen
  \bibfield  {author} {\bibinfo {author} {\bibfnamefont {T.}~\bibnamefont
  {Mori}},\ }\href@noop {} {\bibfield  {journal} {\bibinfo  {journal} {J. Stat.
  Phys.}\ }\textbf {\bibinfo {volume} {159}},\ \bibinfo {pages} {172} (\bibinfo
  {year} {2015}{\natexlab{a}})}\BibitemShut {NoStop}%
\bibitem [{\citenamefont {Hill}(1963)}]{Hill_text}%
  \BibitemOpen
  \bibfield  {author} {\bibinfo {author} {\bibfnamefont {T.~L.}\ \bibnamefont
  {Hill}},\ }\href@noop {} {\emph {\bibinfo {title} {Thermodynamics of small
  systems}}}\ (\bibinfo  {publisher} {Courier Corporation},\ \bibinfo {year}
  {1963})\BibitemShut {NoStop}%
\bibitem [{\citenamefont {Latella}\ \emph {et~al.}(2015)\citenamefont
  {Latella}, \citenamefont {P\'erez-Madrid}, \citenamefont {Campa},
  \citenamefont {Casetti},\ and\ \citenamefont {Ruffo}}]{Latella2015}%
  \BibitemOpen
  \bibfield  {author} {\bibinfo {author} {\bibfnamefont {I.}~\bibnamefont
  {Latella}}, \bibinfo {author} {\bibfnamefont {A.}~\bibnamefont
  {P\'erez-Madrid}}, \bibinfo {author} {\bibfnamefont {A.}~\bibnamefont
  {Campa}}, \bibinfo {author} {\bibfnamefont {L.}~\bibnamefont {Casetti}}, \
  and\ \bibinfo {author} {\bibfnamefont {S.}~\bibnamefont {Ruffo}},\
  }\href@noop {} {\bibfield  {journal} {\bibinfo  {journal} {Phys. Rev. Lett.}\
  }\textbf {\bibinfo {volume} {114}},\ \bibinfo {pages} {230601} (\bibinfo
  {year} {2015})}\BibitemShut {NoStop}%
\bibitem [{\citenamefont {Latella}\ and\ \citenamefont
  {P\'erez-Madrid}(2013)}]{Latella2013}%
  \BibitemOpen
  \bibfield  {author} {\bibinfo {author} {\bibfnamefont {I.}~\bibnamefont
  {Latella}}\ and\ \bibinfo {author} {\bibfnamefont {A.}~\bibnamefont
  {P\'erez-Madrid}},\ }\href@noop {} {\bibfield  {journal} {\bibinfo  {journal}
  {Phys. Rev. E}\ }\textbf {\bibinfo {volume} {88}},\ \bibinfo {pages} {042135}
  (\bibinfo {year} {2013})}\BibitemShut {NoStop}%
\bibitem [{\citenamefont {Mori}(2013{\natexlab{b}})}]{Mori2013_statphys}%
  \BibitemOpen
  \bibfield  {author} {\bibinfo {author} {\bibfnamefont {T.}~\bibnamefont
  {Mori}},\ }\href@noop {} {\bibfield  {journal} {\bibinfo  {journal} {J. Stat.
  Mech.}\ }\textbf {\bibinfo {volume} {2013}},\ \bibinfo {pages} {P10003}
  (\bibinfo {year} {2013}{\natexlab{b}})}\BibitemShut {NoStop}%
\bibitem [{\citenamefont {Mori}(2015{\natexlab{b}})}]{Mori2015_existence}%
  \BibitemOpen
  \bibfield  {author} {\bibinfo {author} {\bibfnamefont {T.}~\bibnamefont
  {Mori}},\ }\href@noop {} {\bibfield  {journal} {\bibinfo  {journal} {J. Phys.
  A: Math. Theor.}\ }\textbf {\bibinfo {volume} {48}},\ \bibinfo {pages}
  {145001} (\bibinfo {year} {2015}{\natexlab{b}})}\BibitemShut {NoStop}%
\bibitem [{\citenamefont {Griffiths}(1968)}]{Griffiths1968}%
  \BibitemOpen
  \bibfield  {author} {\bibinfo {author} {\bibfnamefont {R.~B.}\ \bibnamefont
  {Griffiths}},\ }\href@noop {} {\bibfield  {journal} {\bibinfo  {journal}
  {Phys. Rev.}\ }\textbf {\bibinfo {volume} {176}},\ \bibinfo {pages} {655}
  (\bibinfo {year} {1968})}\BibitemShut {NoStop}%
\bibitem [{\citenamefont {Banerjee}\ \emph {et~al.}(1998)\citenamefont
  {Banerjee}, \citenamefont {Griffiths},\ and\ \citenamefont
  {Widom}}]{Banerjee1998}%
  \BibitemOpen
  \bibfield  {author} {\bibinfo {author} {\bibfnamefont {S.}~\bibnamefont
  {Banerjee}}, \bibinfo {author} {\bibfnamefont {R.}~\bibnamefont {Griffiths}},
  \ and\ \bibinfo {author} {\bibfnamefont {M.}~\bibnamefont {Widom}},\
  }\href@noop {} {\bibfield  {journal} {\bibinfo  {journal} {J. Stat. Phys.}\
  }\textbf {\bibinfo {volume} {93}},\ \bibinfo {pages} {109} (\bibinfo {year}
  {1998})}\BibitemShut {NoStop}%
\bibitem [{\citenamefont {Sauer}(1940)}]{Sauer1940}%
  \BibitemOpen
  \bibfield  {author} {\bibinfo {author} {\bibfnamefont {J.~A.}\ \bibnamefont
  {Sauer}},\ }\href@noop {} {\bibfield  {journal} {\bibinfo  {journal} {Phys.
  Rev.}\ }\textbf {\bibinfo {volume} {57}},\ \bibinfo {pages} {142} (\bibinfo
  {year} {1940})}\BibitemShut {NoStop}%
\bibitem [{\citenamefont {Hiley}\ and\ \citenamefont
  {Joyce}(1965)}]{Hiley-Joyce1965}%
  \BibitemOpen
  \bibfield  {author} {\bibinfo {author} {\bibfnamefont {B.}~\bibnamefont
  {Hiley}}\ and\ \bibinfo {author} {\bibfnamefont {G.}~\bibnamefont {Joyce}},\
  }\href@noop {} {\bibfield  {journal} {\bibinfo  {journal} {Proc. Phys. Soc.}\
  }\textbf {\bibinfo {volume} {85}},\ \bibinfo {pages} {493} (\bibinfo {year}
  {1965})}\BibitemShut {NoStop}%
\bibitem [{\citenamefont {Levy}(1968)}]{Levy1968}%
  \BibitemOpen
  \bibfield  {author} {\bibinfo {author} {\bibfnamefont {P.~M.}\ \bibnamefont
  {Levy}},\ }\href@noop {} {\bibfield  {journal} {\bibinfo  {journal} {Phys.
  Rev.}\ }\textbf {\bibinfo {volume} {170}},\ \bibinfo {pages} {595} (\bibinfo
  {year} {1968})}\BibitemShut {NoStop}%
\bibitem [{\citenamefont {Horner}(1968)}]{Horner1968}%
  \BibitemOpen
  \bibfield  {author} {\bibinfo {author} {\bibfnamefont {H.}~\bibnamefont
  {Horner}},\ }\href@noop {} {\bibfield  {journal} {\bibinfo  {journal} {Phys.
  Rev.}\ }\textbf {\bibinfo {volume} {172}},\ \bibinfo {pages} {535} (\bibinfo
  {year} {1968})}\BibitemShut {NoStop}%
\bibitem [{Note1()}]{Note1}%
  \BibitemOpen
  \bibinfo {note} {Because the shape $\gamma $ can be characterized by a
  function giving the surface of $\gamma $, $Kd\gamma $ can be expressed as
  some functional derivative of the entropy.}\BibitemShut {Stop}%
\bibitem [{\citenamefont {Callen}(1985)}]{Callen_text}%
  \BibitemOpen
  \bibfield  {author} {\bibinfo {author} {\bibfnamefont {H.~B.}\ \bibnamefont
  {Callen}},\ }\href@noop {} {\emph {\bibinfo {title} {Thermodynamics and an
  Introduction to Thermostatistics}}}\ (\bibinfo  {publisher} {Wiley, New
  York},\ \bibinfo {year} {1985})\BibitemShut {NoStop}%
\bibitem [{\citenamefont {Landau}\ and\ \citenamefont
  {Binder}(2000)}]{Landau-Binder_text}%
  \BibitemOpen
  \bibfield  {author} {\bibinfo {author} {\bibfnamefont {D.~P.}\ \bibnamefont
  {Landau}}\ and\ \bibinfo {author} {\bibfnamefont {K.}~\bibnamefont
  {Binder}},\ }\href@noop {} {\emph {\bibinfo {title} {A guide to Monte Carlo
  simulations in statistical physics}}}\ (\bibinfo  {publisher} {Cambridge
  University Press, Cambridge},\ \bibinfo {year} {2000})\BibitemShut {NoStop}%
\bibitem [{\citenamefont {Lahaye}\ \emph {et~al.}(2009)\citenamefont {Lahaye},
  \citenamefont {Menotti}, \citenamefont {Santos}, \citenamefont {Lewenstein},\
  and\ \citenamefont {Pfau}}]{Lahaye2009review}%
  \BibitemOpen
  \bibfield  {author} {\bibinfo {author} {\bibfnamefont {T.}~\bibnamefont
  {Lahaye}}, \bibinfo {author} {\bibfnamefont {C.}~\bibnamefont {Menotti}},
  \bibinfo {author} {\bibfnamefont {L.}~\bibnamefont {Santos}}, \bibinfo
  {author} {\bibfnamefont {M.}~\bibnamefont {Lewenstein}}, \ and\ \bibinfo
  {author} {\bibfnamefont {T.}~\bibnamefont {Pfau}},\ }\href@noop {} {\bibfield
   {journal} {\bibinfo  {journal} {Rep. Prog. Phys.}\ }\textbf {\bibinfo
  {volume} {72}},\ \bibinfo {pages} {126401} (\bibinfo {year}
  {2009})}\BibitemShut {NoStop}%
\bibitem [{\citenamefont {Sandoval-Figueroa}\ and\ \citenamefont
  {Romero-Roch\'{i}n}(2008)}]{Sandoval-Figueroa2008}%
  \BibitemOpen
  \bibfield  {author} {\bibinfo {author} {\bibfnamefont {N.}~\bibnamefont
  {Sandoval-Figueroa}}\ and\ \bibinfo {author} {\bibfnamefont {V.}~\bibnamefont
  {Romero-Roch\'{i}n}},\ }\href {\doibase 10.1103/PhysRevE.78.061129}
  {\bibfield  {journal} {\bibinfo  {journal} {Phys. Rev. E}\ }\textbf {\bibinfo
  {volume} {78}},\ \bibinfo {pages} {061129} (\bibinfo {year}
  {2008})}\BibitemShut {NoStop}%
\end{thebibliography}
\end{document}